\documentclass[conference]{IEEEtran}

\usepackage[authoryear,round]{natbib}
\usepackage[table,xcdraw]{xcolor}
\usepackage{pgfplots}
\pgfplotsset{compat=1.14}
\usepackage{multirow}
\usepackage{courier}
\usepackage{subfig}
\usepackage{amsmath}
\usetikzlibrary{pgfplots.statistics}

\usepackage{float}
\restylefloat{table}
\usepackage{dblfloatfix}

\begin{document}
\IEEEoverridecommandlockouts
\IEEEpubid{\makebox[\columnwidth]{978-1-5090-3050-7/17/\$31.00~\copyright2017 IEEE \hfill} \hspace{\columnsep}\makebox[\columnwidth]{ }}\title{Interpretable Convolutional Neural Networks for Effective Translation Initiation Site Prediction}
%\titleheader{2017 IEEE International Conference on Bioinformatics and Biomedicine (BIBM)}

% \author{\IEEEauthorblockN{Jasper Zuallaert\\ and Mijung Kim\\ and Wesley De Neve}
% \IEEEauthorblockA{Center for Biotech Data Science\\
% Ghent University Global Campus\\
% Songdo, Incheon, 305-701, South Korea\\ 
% IDLab, Ghent University - imec, ELIS\\
% Ghent, 9000, Belgium\\
% Email: jasper.zuallaert@ghent.ac.kr}
% \and
% \IEEEauthorblockN{Yvan Saeys}
% \IEEEauthorblockA{TODO}}
\author{\IEEEauthorblockN{Jasper Zuallaert\IEEEauthorrefmark{1}\IEEEauthorrefmark{2},
Mijung Kim\IEEEauthorrefmark{1}\IEEEauthorrefmark{2},
Yvan Saeys\IEEEauthorrefmark{3}\IEEEauthorrefmark{4}, and 
Wesley De Neve\IEEEauthorrefmark{1}\IEEEauthorrefmark{2}}
\IEEEauthorblockA{\IEEEauthorrefmark{1}Center for Biotech Data Science, Ghent University Global Campus, Songdo, Incheon, 305-701, South Korea}
\IEEEauthorblockA{\IEEEauthorrefmark{2}IDLab, ELIS, Ghent University - imec, Ghent, 9000, Belgium}
\IEEEauthorblockA{\IEEEauthorrefmark{3}VIB-UGent Center for Inflammation Research, Technologiepark 927, 9052 Zwijnaarde-Ghent, Belgium}
\IEEEauthorblockA{\IEEEauthorrefmark{4}Department of Applied Mathematics, Computer Science and Statistics, Ghent University,\\ Krijgslaan 281, S9, 9000 Ghent, Belgium}}

\maketitle

\begin{abstract}

Thanks to rapidly evolving sequencing techniques, the amount of genomic data at our disposal is growing increasingly large. Determining the gene structure is a fundamental requirement to effectively interpret gene function and regulation. An important part in that determination process is the identification of translation initiation sites. In this paper, we propose a novel approach for automatic prediction of translation initiation sites, leveraging convolutional neural networks that allow for automatic feature extraction. Our experimental results demonstrate that we are able to improve the state-of-the-art approaches with a decrease of $75.2\%$ in false positive rate and with a decrease of $24.5\%$ in error rate on chosen datasets. Furthermore, an in-depth analysis of the decision-making process used by our predictive model shows that our neural network implicitly learns biologically relevant features from scratch, without any prior knowledge about the problem at hand, such as the Kozak consensus sequence, the influence of stop and start codons in the sequence and the presence of donor splice site patterns. In summary, our findings yield a better understanding of the internal reasoning of a convolutional neural network when applying such a neural network to genomic data.
\end{abstract}

\IEEEpeerreviewmaketitle

\section{Introduction}

In recent years, with ever-improving sequencing methods, genomic data have become ubiquitous in the field of bioinformatics. For the analysis of gene function and regulation in these data, an overall gene structure is to be determined. This process highly depends on state-of-the-art machine learning approaches to distinguish different regions. In this context, an important task is translation initiation site (TIS) prediction.

A TIS indicates where the translation of genes into proteins starts. In addition to TISs, there are also stop codons (\texttt{TAA}, \texttt{TAG}, or \texttt{TGA}), acting as an endpoint for translation. Other important structures in the translation process are donor splice sites (boundaries between exons and introns) and acceptor splice sites (boundaries between introns and exons). In this paper, we will focus on TIS prediction.

TIS prediction has previously been conducted in a variety of ways. Early methods were mainly based on the consensus sequence \citep{Kozak1987} and the scanning model \citep{Kozak1989}. \cite{Saeys2007a} show that simple models based on these characteristics can already be highly effective. With machine learning techniques becoming more popular, approaches leveraging tools such as Support Vector Machines (SVMs) \citep{Liu2005,Chen2014a} and artificial neural networks \citep{Pedersen1997} have been used effectively.

Recently, the usage of deep convolutional neural networks (CNNs) has become popular for several pattern recognition tasks. Already achieving state-of-the-art results in the domains of image processing \citep{Krizhevsky2012} and natural language understanding \citep{Kim2014}, they are also being applied more and more frequently to genomic data \citep{Alipanahi2015,Kelley2016a,Zeng2016}. A main concern when using CNNs, however, is their black box nature, as the link between input and output may not be evident. Multiple studies have already been conducted on the visualization of the internals of CNNs \citep{Zeiler2014,Shrikumar2017,Fong2017}. 

In this paper, we introduce the usage of CNNs for effective TIS prediction, whilst also investigating the interpretability of these networks. 

%%%%%%%%%%%%%%%%%%%%%% DATA TABLE %%%%%%%%%%%%%%%%%%%%%%%
\begin{table*}[b]
\centering
\renewcommand{\arraystretch}{1.2}
\begin{tabular}{| p{2.75cm}  p{3.5cm} r r l c l |}
\hline
\rowcolor[HTML]{C0C0C0} 
\hline
\textbf{Authors}     & \textbf{dataset}     & \textbf{\# pos} & \textbf{\# neg} & \textbf{pos:neg} & \textbf{seq length} & \textbf{metrics}                          \\
\cite{Saeys2007a} & CCDS (train)  & 13 917            & 350 578          & 1 : 25.2   & 203              & \textit{FPR.80}          \\
&  NCBI for chr. 21 (test)& 258            & 1 267 443          & 1 : 4912.6   &              &           \\
\cite{Chen2014a} & TISdb    & 1159            & 1159         & 1 : 1   & 399              & Se, Sp, Acc, MCC          \\
\hline                    
\end{tabular}
\caption{Characteristics of the datasets used in our experiments.}\label{dataTab}          
\end{table*}
%%%%%%%%%%%%%%%%%%%%%%%%%%%%%%%%%%%%%%%%%%%%%%%%%%%%%%%%%

\section{Methodology}
\label{sec2}
\subsection{Network architecture}
In our approach, we define a CNN-based architecture for TIS prediction, where CNNs are a specialized form of artificial neural networks (ANNs). The latter are able to learn hierarchical feature representations by utilizing multiple connected layers of neurons. Each neuron is responsible for the calculation and propagation of an output signal, using different input signals and corresponding weights. A typical characteristic of CNNs is the presence of convolutional layers, which use a sliding window to detect certain patterns in various positions. In contrast to machine learning approaches that require manual feature definition by human experts, convolutional layers allow for automatic feature extraction from raw data. In our research, DNA sequences are fed to the network using a one-hot vector encoding (i.e., {\tt A} is represented as [1~0~0~0], {\tt C} as [0~1~0~0], and so on).

\begin{figure}
\centering
    	\includegraphics[width=0.42\textwidth]{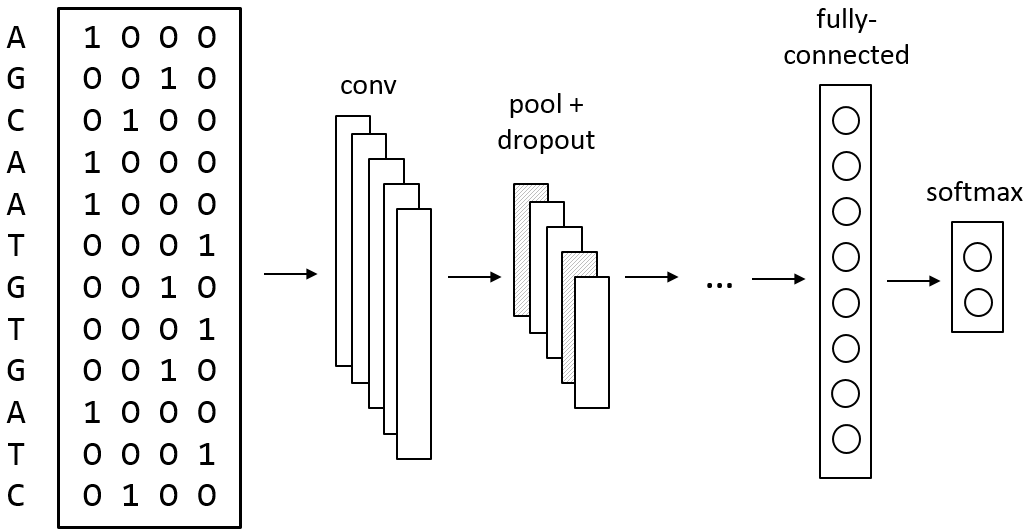}
        \caption{Our network architecture. After converting the sequence in input to a one-hot encoding, we append a number of convolutional, pooling and dropout layers, followed by a fully-connected layer and a softmax layer.}\label{network}
\end{figure}

Our architecture is shown in Fig. \ref{network}. Concretely, it consists of three convolutional layers containing 70, 100, and 150 filters, respectively, of size $7\times4$, $3\times1$, and $3\times1$, respectively. Each of these layers is followed by a max-pooling layer (size $3\times1$ or $4\times1$, depending on the input size) and a dropout layer ($p=0.2$). The output of this last dropout layer is then connected to a fully-connected layer with 512 neurons and another dropout layer ($p=0.2$). Finally, a softmax layer is added, generating a probability that leads to a positive or negative classification of a candidate TIS. In addition, every convolutional and fully-connected layer is succeeded by a rectified linear unit (ReLu) for non-linearization.

Given the aforementioned architecture, we train a predictive model by means of a training set. The weights of the model are initialized using a Glorot uniform distribution, while biases are initialized at zero. The training procedure minimizes the categorical cross-entropy cost function over the training set, using stochastic gradient descent with nesterov momentum. At the start of training, the learning rate is $0.05$, whereafter it is halved every ten epochs, for a total of 50 epochs. After training, the validation set is used to select the predictive model that performed best during training. Finally, we evaluate the selected model against a testing set.

\subsection{Visualization}
After evaluation, we want to interpret the choices made by our predictive model. To that end, we utilize the state-of-the-art DeepLIFT algorithm, as described in \cite{Shrikumar2017}. This algorithm generates contribution scores for each nucleotide in the input sequence, indicating how important they are in determining the final prediction. Using these contribution scores, we can then investigate which nucleotide patterns the network is most sensitive to.

Before determining the nucleotide patterns that trigger a strong network response, we adjust the contribution scores of every individual input sequence to act on the same scale. To that end, we use the following formula to acquire weighted contribution scores ({\tt wcs}), which indicate the percentage of how much each individual nucleotide contributes to the prediction, as compared to the other nucleotides:

\begin{equation}
wcs_i = \frac{cs_i}{\displaystyle\sum_{j=1}^{n} |cs_j|},
\end{equation}
with $wcs_i$ denoting the weighted contribution score at position $i$, with $cs_i$ denoting the contribution score for the nucleotide at position $i$, and with $n$ denoting the sequence length.

Next, using the weighted contribution scores, we identify which patterns are important in the TIS prediction conducted by the network. The visualization of these results can be found in Section \ref{sec3}.

\subsection{Data and metrics}

To evaluate our approach, we compare our predictive model with two different state-of-the-art techniques. The datasets used are detailed in Table \ref{dataTab}, accompanied by the metrics used. The sensitivity (\textit{Se}), specificity (\textit{Sp}), false positive rate (\textit{FPR}), accuracy (\textit{Acc}), and Mathew's correlation coefficient (\textit{MCC}) are defined as follows:

\[Se = \dfrac{TP}{TP+FN} \hspace{0.4cm} Sp = \dfrac{TN}{TN+FP} \hspace{0.4cm} FPR = \dfrac{FP}{TN+FP} \]
\[Acc = \dfrac{TN+TP}{TN+TP+FP+FN}\]
\[MCC = \dfrac{TP*TN - FP*FN}{\sqrt[]{(TP+FP)(TP+FN)(TN+FP)(TN+FN)}}\]
where $TP$ denotes the number of true positives, $TN$ the number of true negatives, $FP$ the number of false positives, and $FN$ the number of false negatives. 

We will also make use of the notation \textit{FPR.80}, which denotes the \textit{FPR} for a fixed \textit{Se} of $0.80$. 

\section{Results and Discussion}
\label{sec3}

\subsection{Comparison with other approaches}
In \cite{Saeys2007a}, the authors use a combination of three simple TIS prediction mechanisms: the position weight matrix, stop codon frequencies, and an interpolated context model to investigate the composition of the regions around the investigated TIS. Evaluation is conducted on human chromosome 21, extracted from NCBI build 36 version 1, after training on the CCDS (consensus CDS) dataset (excluding chromosome 21).

\begin{table}[H]
  \centering
  \begin{tabular}{c|c}
    approach & \textit{FPR.80} \\
    \hline
    \cite{Saeys2007a} & 0.125\\
    Our approach & \textbf{0.031}\\
  \end{tabular}
  \vspace{0.3cm}
  \caption{Results obtained for the NCBI chromosome 21 dataset, after training on the CCDS dataset.}\label{compSaeys}
\end{table}

Table \ref{compSaeys} shows the results obtained. Our model was able to classify the candidate TISs more effectively, decreasing the $FPR.80$ from $0.125$ to $0.031$, resulting in a relative decrease of $75.2\%$. This performance increase is also shown in Figure~\ref{fprtpr}, where we show the ROC curves for both models.

This result encourages the usage of CNNs for DNA analysis. Convolutional layers are very suited for the recognition of different patterns. When searching for decisive patterns, the first convolutional layer gives every nucleotide an individual weight within that pattern, yielding an activation score at every position. By reducing the dimensionality of the activations as we propagate through the network, those activations then get combined to achieve a high-level evaluation of the input.

Additionally, we compare our approach to the approach used in \cite{Chen2014a}, which uses SVMs in combination with manually defined feature vectors, taking into account nucleotide frequencies whilst retaining considerable sequence order information, but also leveraging physicochemical properties of codons. In this case, evaluation is conducted on a much smaller dataset (see Table \ref{dataTab}), by making use of leave-one-out cross-validation.

\pgfplotstableread{data/curve_my_data.dat}{\curveMyData}
\pgfplotstableread{data/curve_yvan_data.dat}{\curveYvanData}
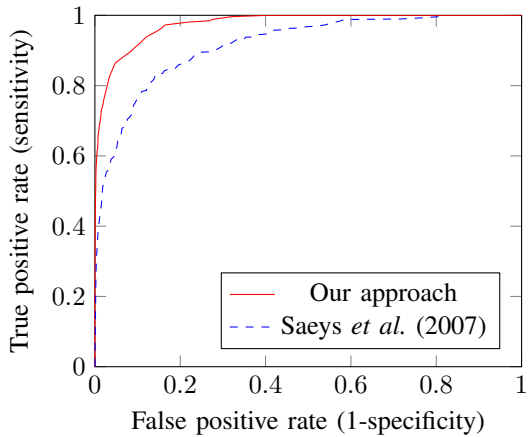
\begin{figure}
\centering
\begin{tikzpicture}
        \begin{axis}[
        		xlabel = {False positive rate (1-specificity)},
                ylabel = {True positive rate (sensitivity)},
                enlarge x limits=0,
                enlarge y limits=0, 
                width = 0.4\textwidth,
            	legend style={at={(0.95,0.05)},anchor=south east},
                  ]
              \addplot [mark=none,color=red] table [y=tpr,x=fpr] {\curveMyData};
              \addlegendentry{Our approach}
              \addplot [mark=none,color=blue,dashed] table [y=tpr,x=fpr] {\curveYvanData};
              \addlegendentry{Saeys \textit{et al.} (2007)}
        \end{axis}
\end{tikzpicture}\caption{ROC curves, plotting the relation between true and false positive rates, when classifying using different decision thresholds.}\label{fprtpr}
\end{figure}

\begin{table}[H]
  \centering
  \begin{tabular}{c|cccc}
    approach & \textit{Se} & \textit{Sp} & \textit{Acc} & \textit{MCC}  \\
    \hline
    \cite{Saeys2007a} & 0.9532 & 0.9643 & 0.9602 & 0.921 \\
    \cite{Chen2014a} & 0.9749 & \textbf{0.9842} & 0.9792 & 0.958 \\
    Our approach & \textbf{0.9905} & 0.9783 & \textbf{0.9843} & \textbf{0.969}\\
  \end{tabular}
  \vspace{0.3cm}
  \caption{Results obtained for the cross validation tests executed on the TISdb dataset.}\label{compChen}
\end{table}

Results are listed in Table \ref{compChen}. It shows that our approach, which does not need manually defined features, is more effective. Whereas \cite{Chen2014a} achieves an accuracy of $0.9792$, our approach reaches an accuracy of $0.9838$, leading to a relative error rate reduction of $24.5\%$.

\subsection{Visualization}
We visualize the model acquired for our experiments using the CCDS dataset. First, we calculate the {\tt wcs} (as described in Section \ref{sec2}) for every input sequence in the NCBI chromosome 21 dataset. Using these scores, we can then calculate the averages on each position for each nucleotide or pattern. That way, we are able to verify four hypotheses:

\textbf{\begin{enumerate}\addtocounter{enumi}{0}
\item In the proximity of a candidate TIS, the network looks for a pattern that resembles the well-known Kozak consensus sequence \citep{Kozak1987}.
\end{enumerate}}
In Fig.~\ref{motifs}a, we show the average {\tt wcs} per nucleotide on each position in the proximity of the splice site. This is visualized by having higher bars for higher scores. As a comparison, we also show the aforementioned consensus sequence in Fig.~\ref{motifs}b, which is based on the frequencies at each position (a higher frequency results in a bigger letter), over a known set of TISs \citep{Cavener1987}. When comparing both visualizations, we can observe that, for each of the nine positions in the region $[-6,-1]+[3,5]$, the nucleotide with the highest frequency is the nucleotide that gets the highest average {\tt wcs}. 

\textbf{\begin{enumerate}\addtocounter{enumi}{1}
\item The network learns that genes are composed of nucleotide triplets (codons). In addition, the network is highly sensitive to the three stop codons \texttt{TAA}, \texttt{TAG}, and \texttt{TGA}. If a stop codon occurs behind the TIS, the chances of a false classification increase, as this would imply an unlikely short gene.
\end{enumerate}}

Fig.~\ref{TAAgraph} shows the average {\tt wcs} per position for the stop codon pattern {\tt TAA}. The spikes in the graph occur every three positions. This behavior can be attributed to the translation mechanism, where RNA is translated in codons. The network only steers towards a negative classification when the \texttt{TAA} pattern would indeed get translated into a protein, not if the pattern is split over two codons. 

In Fig.~\ref{stopcodons}, we present the average scores per stop codon, as well as the averages for the \texttt{ATG} pattern and the averages over all patterns (\texttt{NNN}). Clearly, the presence of any stop codon behind the TIS that would get translated, has a negative impact on the prediction outcome.

\textbf{\begin{enumerate}\addtocounter{enumi}{2}
\item The network is sensitive to the occurrence of another \texttt{ATG} pattern. This is justified by the Scanning Model \citep{Kozak1989}, which implies that the 40S ribosomal subunit, responsible for the initiation of translation, generally stops at the first \texttt{AUG} (\texttt{ATG} in DNA) it encounters after binding to the 5'-end of mRNA.
\end{enumerate}}

As seen in Fig.~\ref{stopcodons}, the scoring for occurrences of the pattern \texttt{ATG} shows a similar behavior, although inversed: the network is negatively biased regarding \texttt{ATG} codons in front of the TIS, rather than behind it, which confirms that the Scanning Model is implicitly learnt by the network.

\textbf{\begin{enumerate}\addtocounter{enumi}{3}
\item The network is sensitive to donor splice sites. When it notices a typical donor splice site pattern behind the TIS, the chance of a positive classification increases, as this suggests the presence of an exon. 
\end{enumerate}}
Fig.~\ref{MAGGTAAGgraphL} shows the average {\tt wcs} for each position of the pattern \texttt{MAG\textbf{GT}AAGT} (with \texttt{M} = \texttt{A} or \texttt{M} = \texttt{C}), which is the typical consensus sequence for donor splice sites \citep{Iwata2011}. The {\tt wcs} values in front of the splice site vary between $-6$ and $-14$, and behind the splice site between $10$ and $20$. Consequently, we can conclude that the network learns that a donor splice site pattern should generally only occur behind the splice site.

%%%%%%%%%%%%%%%% FIGURES %%%%%%%%%%%%%%%
\setlength{\abovecaptionskip}{0pt}
\setlength{\belowcaptionskip}{-2ex}
\pgfplotstableread{data/stop_codons_0.dat}{\stopcodonsNul}
\pgfplotstableread{data/stop_codons_1.dat}{\stopcodonsEen}
\pgfplotstableread{data/stop_codons_2.dat}{\stopcodonsTwee}
\pgfplotstableread{data/graph_TAA.dat}{\TAAgraph}
\pgfplotstableread{data/graph_MAGGTAAG.dat}{\MAGGTAAGgraph}

\pgfplotstableread{data/kozak_data.dat}{\kozakData}

%%%% KOZAK SEQUENCE %%%%
\definecolor{green}{rgb}{0.0,0.62,0.33}
\definecolor{forange}{rgb}{0.99, 0.75, 0.0}
\definecolor{blue}{rgb}{0.25, 0.41, 0.82}
\begin{figure*}
	\begin{tabular}{cc}
    	\begin{tikzpicture}
            \begin{axis}[
            ybar stacked,
            bar width = 11pt,
            width=0.49\textwidth,
            height=0.32\textwidth,
            xlabel=Position,
            ylabel=average wcs,
            ymax = 6,
            tick pos = left,
            extra y ticks = 0,
            extra x ticks = {-6,-5,-4,-3,-2,-1,0,1,2,3,4,5},
            extra y tick labels = ,
            extra y tick style  = { grid = major },
            legend columns = 4,
            legend style={at={(0.99,0.99)},anchor=north east},
            xticklabel style={rotate=90},
            x label style={at={(0.5,-0.085)}}
            ]
            \addplot[color=green,fill=green] table [y=A,x=Pos] {\kozakData};
            \addlegendentry{A}
            \addplot[color=blue,fill=blue] table [y=C,x=Pos] {\kozakData};
            \addlegendentry{C}
            \addplot[color=forange,fill=forange] table [y=G,x=Pos] {\kozakData};
            \addlegendentry{G}
            \addplot[color=red,fill=red] table [y=T,x=Pos] {\kozakData};
            \addlegendentry{T}
            \end{axis}
        \end{tikzpicture}
    &
    \includegraphics[width=0.49\textwidth]{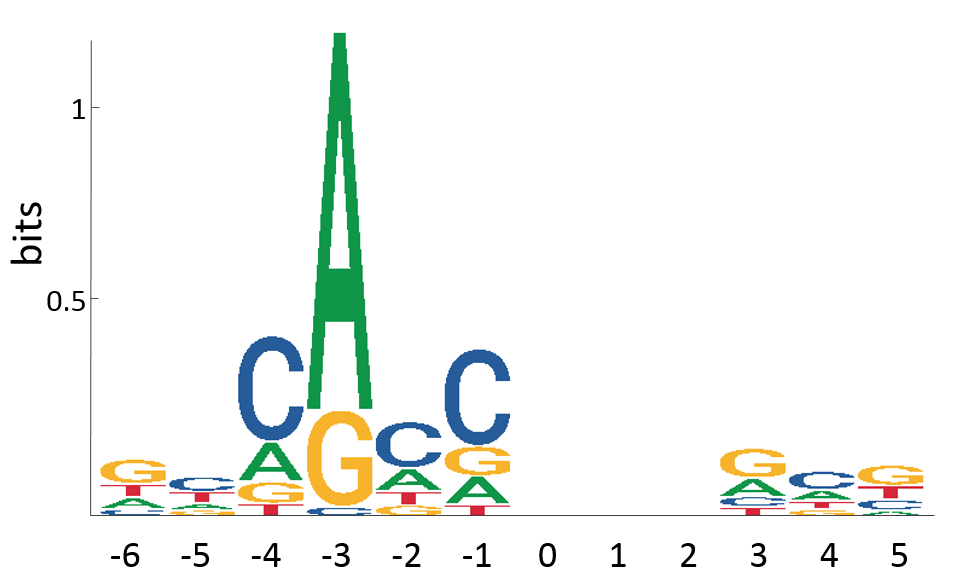}
    	\end{tabular}
    \caption{The average {\tt wcs} for each nucleotide in the proximity of the TIS, per position (left). For comparison purposes, we also show the Kozak consensus sequence (right).}
    \label{motifs}
\end{figure*}
%%%% TAA GRAPH %%%%
\begin{figure*}
        \begin{tikzpicture}
        \begin{axis}[width=1.0\textwidth,height=0.25\textwidth,tick pos=left,
        			xlabel={Position},ylabel={average wcs},
                    enlarge x limits=0.00,
                    extra y ticks = 0,
                    extra y tick labels = ,
                    extra y tick style  = { grid = major }]
            \addplot[color=red,unbounded coords=jump] table {\TAAgraph};
        \end{axis}
        \end{tikzpicture}
    \caption{The average {\tt wcs} of the pattern \texttt{TAA} when appearing at the indicated starting positions.}\label{TAAgraph}
\end{figure*}
%%%% ATG TAA TAG TGA NNN %%%%
\begin{figure*}
	\begin{tabular}{ccc}
        \begin{tikzpicture}
        \begin{axis}[
                  height=0.3*\textwidth,
                  major x tick style = transparent,
                ybar=2*\pgflinewidth,
                bar width=8 pt,
                ymajorgrids = true,
                ylabel = {average wcs},
                symbolic x coords={ATG,TAA,TAG,TGA,NNN},
                xtick = data,
                scaled y ticks = false,
                enlarge x limits=0.15,
                ymin=-20,
                ymax=10
                  ]
              \addplot table [y=before,x=Label] {\stopcodonsNul};
              \addplot table [y=after,x=Label] {\stopcodonsNul};
        \end{axis}
        \end{tikzpicture}
        & 
        \begin{tikzpicture}
        \begin{axis}[
                  height=0.3*\textwidth,
                  major x tick style = transparent,
                ybar=2*\pgflinewidth,
                bar width=8 pt,
                ymajorgrids = true,
                symbolic x coords={ATG,TAA,TAG,TGA,NNN},
                xtick = data,
                scaled y ticks = false,
                enlarge x limits=0.15,
                legend style={at={(0.92,0.03)},anchor=south east,column sep=1ex},
                ymin=-20,
                ymax=10
                  ]
              \addplot table [y=before,x=Label] {\stopcodonsEen};
              \addplot table [y=after,x=Label] {\stopcodonsEen};
              \legend{region in front of TIS, region behind TIS}
        \end{axis}
        \end{tikzpicture}
        &
        \begin{tikzpicture}
        \begin{axis}[
                  height=0.3*\textwidth,
                  major x tick style = transparent,
                ybar=2*\pgflinewidth,
                bar width=8 pt,
                ymajorgrids = true,
                symbolic x coords={ATG,TAA,TAG,TGA,NNN},
                xtick = data,
                scaled y ticks = false,
                enlarge x limits=0.15,
                ymin=-20,
                ymax=10
                  ]
              \addplot table [y=before,x=Label] {\stopcodonsTwee};
              \addplot table [y=after,x=Label] {\stopcodonsTwee};
        \end{axis}
        \end{tikzpicture}
        \\ (a) start at codon position 0 & (b) start at codon position 1 & (c) start at codon position 2
    \end{tabular} 
    \caption{The average {\tt wcs} scores for the listed triplets, when occurring in the 60 positions in front of (blue) or behind (red) the TIS. The averages are split up in three groups: (a) the pattern is aligned properly, meaning that it would be translated as a triplet during translation (e.g., in \texttt{ATG \textbf{GAT} GCG}, the \texttt{GAT} would be translated as a codon), (b) the pattern is aligned to be starting at the second position within a codon (e.g., if \texttt{ATG G\textbf{AT G}CG} would be translated, the indicated \texttt{ATG} starts at the second position within that codon), and (c) the pattern is aligned to be starting at the last position within a codon (e.g., \texttt{ATG GA\textbf{T GC}G}).}\label{stopcodons}
\end{figure*}
%%%% MAGGTAAG %%%%
\begin{figure*}
        \begin{tikzpicture}
        \begin{axis}[width=1.0\textwidth,height=0.25\textwidth,tick pos=left,
        			xlabel={Position},ylabel={average wcs},
                    enlarge x limits=0.05,
                    extra y ticks = 0,
                    extra y tick labels = ,
                    extra y tick style  = { grid = major }]
            \addplot[color=red,unbounded coords=jump] table {\MAGGTAAGgraph};
        \end{axis}
        \end{tikzpicture}
   \caption{The average {\tt wcs} of the pattern \texttt{MAGGTAAGT} when appearing at the indicated starting positions.}\label{MAGGTAAGgraphL}
\end{figure*}

%%%%%%%%%%%%%%%% CONCLUSION %%%%%%%%%%%%%%%

\section{Conclusions and Future Work}
In this paper, we successfully applied convolutional neural networks to the problem of translation initiation site prediction. When comparing to the state-of-the-art, we were able to reduce the false positive rate for a sensitivity of $0.80$ by $75.2\%$ in a first experiment, and the classification error rate by $24.5\%$ in a second experiment, with both experiments conducted on different datasets and compared to different approaches.

Furthermore, we were able to provide insight into the decisions taken by our network. By calculating the contribution scores of the individual nucleotides towards the final prediction, and by subsequently looking for reoccurring structures in these scores, we were able to verify four hypotheses. In particular, we found that our network learns the typical pattern for a TIS, that it is sensitive towards stop codons behind and other start codons in front of the TIS, and that it is negatively biased towards donor site patterns preceding the TIS and positively biased towards those succeeding it.

We believe the aforementioned findings are of high interest, as they prove that our network implicitly learns biologically relevant features, without having any prior knowledge about the problem. All these features are automatically learnt from scratch, based on a dataset consisting of raw DNA sequences with corresponding labels. In the future, leveraging the kind of visualizations presented in this paper could open the door for the discovery of new biological characteristics in various use-cases.

\section*{Funding}
The research activities described in this paper were funded by Ghent University Global Campus, Ghent University, imec, Flanders Innovation \& Entrepreneurship (VLAIO), the Fund for Scientific Research-Flanders (FWO-Flanders), and the EU. Yvan Saeys is a Marylou Ingram Scholar.

\bibliographystyle{natbib}
\bibliography{IEEE_paper}

\end{document}